\documentclass[%
reprint, twocolumn,
superscriptaddress,
nofootinbib,
amsmath,amssymb, aps, pra,
]{revtex4}

\usepackage{amsmath}
\usepackage{amssymb}
\usepackage{graphicx}
\usepackage{dcolumn}
\usepackage{bm}
\usepackage{color} 
\usepackage{CJK}
\usepackage{graphicx}
\usepackage{subfigure}
 \usepackage{mathrsfs}
 \usepackage{dsfont}
\usepackage[extension=xxx]{hyperref}

\def\la{{\langle}}
\def\ra{{\rangle}}

\newcommand{\beq}{\begin{equation}}
\newcommand{\eeq}{\end{equation}}
\newcommand{\beqa}{\begin{eqnarray}}
\newcommand{\eeqa}{\end{eqnarray}}

\begin{document}
\title{Hamiltonian design to prepare arbitrary states of four-level systems}
\author{Y.- C. Li}
\affiliation{Department of Physics, Shanghai University, 200444 Shanghai, People's Republic of China}
\author{D. Mart\'\i nez}
\affiliation{Department of Physical Chemistry, Universidad del Pa\'{\i}s Vasco - Euskal Herriko Unibertsitatea,
Apdo. 644, Bilbao, Spain}
\author{S. Mart\'\i nez-Garaot}
\affiliation{Department of Physical Chemistry, Universidad del Pa\'{\i}s Vasco - Euskal Herriko Unibertsitatea,
Apdo. 644, Bilbao, Spain}
\author{X. Chen}
\affiliation{Department of Physics, Shanghai University, 200444 Shanghai, People's Republic of China}
\author{J. G. Muga}
\affiliation{Department of Physical Chemistry, Universidad del Pa\'{\i}s Vasco - Euskal Herriko Unibertsitatea,
Apdo. 644, Bilbao, Spain}
\begin{abstract}
We propose a method to manipulate, possibly faster than adiabatically,  four-level systems with time-dependent couplings and constant
energy shifts (detunings in quantum-optical realizations).
We inversely engineer the Hamiltonian, in ladder, tripod, or diamond configurations,  to prepare arbitrary states
using the geometry of four-dimensional rotations to set the state populations,
specifically we use Cayley's factorization of a general rotation into  right- and left-isoclinic rotations.
\end{abstract}
\maketitle
\section{Introduction}
The coherent state manipulation and control of multiple-level quantum systems plays a significant role in atomic,
molecular and optical  physics, with applications in existing or developing quantum  technologies and quantum information processing \cite{Mikio}.
Slow adiabatic protocols may be used but they require  long times, and detrimental effects of noise and perturbations
accumulate.  This has motivated  the development of  a set of techniques denominated
``shortcuts to adiabaticity" to speed up the processes, which include
counter-diabatic driving \cite{Berry,Rice}, inverse engineering based on invariants \cite{Xi1},
Lie algebraic methods \cite{MGLie,ELie},
fast quasi-adiabatic approaches \cite{FAQUAD},
or fast-forward approaches \cite{ff1,ff2,ff3}.

Some of these methods require  to add terms in the Hamiltonian which are not easy or possible to implement
in practice \cite{lyc,tr3,tr6,Xi1}.
This problem has been addressed in specific systems by optimizing physically available terms \cite{tr6}, or by unitary transformations making use of the Lie algebraic structure  of the dynamics \cite{tr1,tr2,tr3,tr4,tr5}. However, generic solutions are not known and, as the system complexity
and number of generators increase, the Lie algebraic methods may become numerically unstable or
cumbersome to apply.
These difficulties may be already noticed in three-level  or four-level systems,
so  alternative or complementary approaches are currently being explored.

In Ref. \cite{Rodrigue} the authors proposed a  scheme  to control three-level system dynamics by separating the evolution into
population changes, which may be parameterized using Rodrigues' rotation formula, and phase changes. This separation was used to inversely
construct the Hamiltonian of the three-level system so as to drive a given transition with allowed couplings and vanishing forbidden couplings.
Our goal here is to explore the extension of this concept to four-level systems. Certain couplings should not appear in the final Hamiltonian
to implement specific 4-level configurations such as a ``diamond'', a ``tripod'', or a ``ladder''.
The population dynamics is now represented by rotations in a four dimensional space, which are considerably more complex
and less intuitive than in three dimensions. We have found a description of the rotation in terms of isoclinic
matrices and quaternions, making use of Cayley's factorization,  more convenient to perform the inversion than a generalized Rodrigues' formula,
see Sec. \ref{section1}. In Sec. \ref{section2} we find the Hamiltonian for the different configurations and provide examples. The appendices address technical points: long formulae in Appendix A, a short account of quaternions for 4D rotations in Appendix B, 
and details of quantum optical realizations in Appendix C.

Four-level systems are widely found and used in different contexts such as atomic physics, optical lattices \cite{N1Xu,N2Goren,D2Thomas} or waveguides \cite{tri4Vitanov1,tri5Vitanov2,tri6Ichimura}, with applications such as  electromagnetically induced transparency (EIT) \cite{N1Xu,tri2Corbalan,D4Bao-Quan Ou}, electromagnetically induced absorption \cite{N2Goren}, or beam splitting \cite{tri4Vitanov1,tri5Vitanov2}.
Most of the results in this paper are set in an abstract way, without specifying necessarily the physical system,
but the notation is chosen as in a quantum-optical realization where atomic internal levels are coupled by
laser fields, consistent with Rabi frequencies or detunings as matrix elements of the Hamiltonian.
An explicit connection for the diamond configuration is worked out in
Appendix C.
\section{4D rotations}
\label{section1}
Consider a four-level system in the state $|\psi(t)\ra=c_1(t)|1\ra+c_2(t)e^{i\varphi_2(t)}|2\ra+c_3(t)e^{i\varphi_3(t)}|3\ra+c_4(t)e^{i\varphi_4(t)}|4\ra$,
where $c_n(t)$ are real probability amplitudes of bare states $|n\ra$ satisfiying the normalization
 $c_1^2(t)+c_2^2(t)+c_3^2(t)+c_4^2(t)=1$,
and the $\varphi_n(t)$ are relative phases. Following \cite{Rodrigue}, we separate phase and amplitude
information by writing  $|\psi(t)\ra=K(t)|\psi_r(t)\ra$, where $K(t)=|1\ra \la 1|+e^{i\varphi_2(t)}|2\ra \la 2|+e^{i\varphi_3(t)}|3\ra \la 3|+e^{i\varphi_4(t)}|4\ra \la 4|$ and $|\psi_r(t)\ra=c_1(t)|1\ra+c_2(t)|2\ra+c_3(t)|3\ra+c_4(t)|4\ra$. $K(t)$ is a unitary transformation that contains the phases and
$|\psi_r(t)\ra$ represents a 4-dimensional (4D) vector on the surface of a 4D sphere. The states $|\psi(t)\ra$
and $|\psi_r(t)\ra$
evolve via time-evolution operators $U(t)$ and $U_r(t)$ related by $U_r(t)=K^\dagger(t)U(t)K(0)$,
\beqa
|\psi(t)\ra&=&U(t)|\psi(0)\ra,
\nonumber \\
|\psi_r(t)\ra&=&U_r(t)|\psi_r(0)\ra,
\eeqa
where we set initial time as $0$.
$U_r(t)$ represents
a 4D rotation displacing points on the surface of the 4D sphere.
In the four-dimension real space, we define the
rotation Hamiltonian as
\beq
\label{Hr1}
H_r(t)=i \hbar \dot{U}_r(t) U_r^\dagger (t),
\eeq
such that $i \hbar \dot{U}_r(t)= H_r(t) U_r(t)$, whereas  the total Hamiltonian is
\beqa
\label{H1}
H(t)&=&i \hbar \dot{U}(t) U^\dagger (t)
\nonumber\\
&=&i \hbar \dot{K}(t)K^\dagger (t)+K(t)H_r(t)K^\dagger(t).
\eeqa
\subsection{Rotations in $\mathbb{E}^4$}
\label{subsectionA}
In four dimensional Euclidean space $\mathbb{E}^4$, a 4D rotation with centre $O$
can be expressed
by a rotation matrix \cite{ro1Erdogdu,ro2Jason,ro3Cole}
\beq
\begin{pmatrix}
\label{matrix1}
\cos\alpha&-\sin\alpha&0&0\\
\sin\alpha&\cos\alpha&0&0\\
0&0&\cos\beta&-\sin\beta\\
0&0&\sin\beta&\cos\beta\\
\end{pmatrix}
\eeq
in some appropriate orthogonal coordinates $\tilde{w}\tilde{x}\tilde{y}\tilde{z}$.
Instead of having an axis of rotation as in 3D,  4D rotations are defined by a pair of
completely orthogonal planes of rotation ($\tilde{w}$-$\tilde{x}$ and $\tilde{y}$-$\tilde{z}$ in the example),
$\alpha$ and $\beta$ are the angles of rotation with respect to the origin of any point on the  $\tilde{w}$-$\tilde{x}$ and $\tilde{y}$-$\tilde{z}$ planes, respectively.
More details can be found e.g. in \cite{ro1Erdogdu,ro2Jason,ro3Cole}.

We may classify the rotations based on the $\alpha$ and $\beta$ angles:

If $\alpha\neq\beta\neq0$, the rotation is a \emph{double rotation}. There are two completely orthogonal (invariant) planes of rotation, with just the point $O$
in common. Points in the first plane rotate through $\alpha$ with respect to the origin, and in the second plane rotate through $\beta$. For a general double rotation the planes of rotation and angles are unique.
Points which are not in the two planes rotate with respect to the origin through an angle between $\alpha$ and $\beta$.

If either of $\alpha$ or $\beta$ are zero, the rotation is a \emph{simple rotation}
about the rotation center $O$: There is a fixed plane whose points do not change,
whereas  half-lines from $O$ orthogonal to this plane are displaced through the non-zero angle ($\alpha$ or $\beta$).

If $\alpha=\pm \beta$ the rotation is {\it isoclinic} and all non-zero points are rotated through the same angle.
Then there are infinitely many pairs of orthogonal planes that can be treated as planes of rotation \cite{ro1Erdogdu}.
An isoclinic rotation can be left- or right-isoclinic (depending on whether $\alpha$=$\beta$ or $\alpha$=-$\beta$) \cite{iso1Federico}. According to Cayley's factorization \cite{iso2Thomas,iso3Cayley}, any 4D rotation matrix can be decomposed into the product of a right- and a left-isoclinic matrix.
This decomposition is also conveniently expressed in terms of quaternions, as discussed in the following subsection.\vspace*{.4cm}
\subsection{Isoclinic rotations and quaternions}
In 4D Euclidean space, an arbitrary point $C$ can be represented as a column vector $(w,x,y,z)^T$
or as  $C=w+x \textbf{\textit{i}}+y \textbf{\textit{j}}+z \textbf{\textit{k}}$ \cite{qua2Mebius1,qua3Mebius2}. If $|C|^2=w^2+x^2+y^2+z^2=1$
we call it unit quaternion. A general 4D rotation takes $C$ to $C'$, according to
\beq
\label{qp}
C'=q C p,
\eeq
where $q=q_w+q_x \textbf{\textit{i}}+ q_y \textbf{\textit{j}}+ q_z \textbf{\textit{k}}$ and $p=p_w+ p_x \textbf{\textit{i}}+ p_y \textbf{\textit{j}}+ p_z\textbf{\textit{k}}$ are two unit quaternions. See the Appendix A for a minimal introduction to quaternion algebra.
In more common matrix language, the rotation reads
\begin{widetext}
\beq
\label{rotationsym}
C'=M_L M_R C,
\eeq
\beq
\label{rotationmatrix}
\begin{pmatrix}
w'\\
x'\\
y'\\
z'\\
\end{pmatrix}=
\begin{pmatrix}
q_w&-q_x&-q_y&-q_z\\
q_x&q_w&-q_z&q_y\\
q_y&q_z&q_w&-q_x\\
q_z&-q_y&q_x&q_w\\
\end{pmatrix}
\begin{pmatrix}
p_w&-p_x&-p_y&-p_z\\
p_x&p_w&p_z&-p_y\\
p_y&-p_z&p_w&p_x\\
p_z&p_y&-p_x&p_w\\
\end{pmatrix}
\begin{pmatrix}
w\\
x\\
y\\
z\\
\end{pmatrix},
\eeq
\end{widetext}
a formula due to Van Elfrinkhof \cite{iso1Federico,iso2Thomas}.  $M_L$ and $M_R$ are isoclinic
matrices \cite{qua2Mebius1,qua3Mebius2}, so $R=M_L M_R=M_R M_L$ is a 4D rotation matrix without loss of generality. Furthermore,  $R^\dagger R=R R^\dagger=I$ due to $|q|^2=q_w^2+q_x^2+q_y^2+q_z^2=1$ and $|p|^2=p_w^2+p_x^2+p_y^2+p_z^2=1$. A summary of further relations between quaternions and 4D-rotations, such as the relation between the isoclinic matrices and the orthogonal rotation planes and corresponding rotation angles,  may be found in Appendix A.
\section{Hamiltonian inverse engineering\label{section2}}
In this section, we will make use of the rotation formula (\ref{rotationmatrix})
to engineer the Hamiltonian and dynamics to drive a four-level system from an initial state to a final state. We substitute $U_r(t)=R(t)$ in Eq. (\ref{Hr1}), where the quaternion components are generally time dependent.
The corresponding rotation Hamiltonian has the following structure
\beqa
\label{Hr2}
H_r(t)&=&i \hbar \dot{U}_r(t) U_r^\dagger (t) \nonumber
\\
&=&i \hbar
\begin{pmatrix}
0&\Omega_{12}(t)&\Omega_{13}(t)&\Omega_{14}(t)
\\
-\Omega_{12}(t)&0&\Omega_{23}(t)&\Omega_{24}(t)
\\
-\Omega_{13}(t)&-\Omega_{23}(t)&0&\Omega_{34}(t)
\\
-\Omega_{14}(t)&-\Omega_{24}(t)&-\Omega_{34}(t)&0
\end{pmatrix}, \nonumber \\
\eeqa
where the real elements $\Omega_{nm}(t)$ are functions of the unit quaternion components (the explicit expression
is given in  Appendix B).

\begin{figure}[t]
\scalebox{0.5}[0.5]{\includegraphics{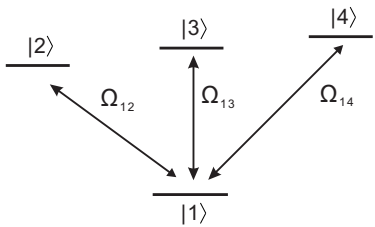}}
\caption{Energy level scheme for the inverse-tripod configuration with three
non-zero couplings  $\Omega_{12}$, $\Omega_{13}$ and $\Omega_{14}$.}
\label{config2}
\end{figure}

Taking the relative phases into account,  the total Hamiltonian (\ref{H1}) is
\beqa
\label{H2}
H(t)&=&i \hbar \dot{U}(t) U^\dagger (t)
\nonumber\\
&=&i \hbar \dot{K}(t)K^\dagger (t)+K(t)H_r(t)K^\dagger(t)
\nonumber\\
&=& \hbar [-\dot{\varphi}_2(t)|2\ra \la 2|-\dot{\varphi}_3(t)|3\ra \la 3|-\dot{\varphi}_4(t)|4\ra \la 4|
\nonumber\\
&+&i(e^{-i \varphi_2(t)}\Omega_{12}(t)|1\ra \la 2|+e^{-i \varphi_3(t)} \Omega_{13}(t)|1\ra \la 3|
\nonumber\\
&+&e^{-i \varphi_4(t)}\Omega_{14}(t)|1\ra \la 4|+e^{i [\varphi_2(t)-\varphi_3(t)])}\Omega_{23}(t)|2\ra \la 3|
\nonumber\\
&+&e^{i [\varphi_2(t)-\varphi_4(t)]}\Omega_{24}(t)|2 \ra \la 4|
\nonumber\\
&+&e^{i [\varphi_3(t)-\varphi_4(t)]}\Omega_{34}(t)|3\ra \la 4|)]+ H.c.
\eeqa
The physical interpretation of this Hamiltonian depends on the system considered.
In quantum optics this is to be interpreted as  an interaction picture Hamiltonian where the diagonal terms are not energies of the bare levels, as depicted e.g. in Fig. \ref{config2}, but detunings, see Appendix C.

It proves useful to parameterize the quaternion components in terms of generalized
spherical angles \cite{Sommer,MW},
\beqa
\label{parametrization}
q_w&=&\cos\gamma_1,
\nonumber\\
q_x&=&\sin\gamma_1 \cos\theta_1,
\nonumber\\
q_y&=&\sin\gamma_1 \sin\theta_1 \cos\phi_1,
\nonumber\\
q_z&=&\sin\gamma_1 \sin\theta_1 \sin\phi_1,
\nonumber\\
p_w&=&\cos\gamma_2,
\nonumber\\
p_x&=&\sin\gamma_2 \cos\theta_2,
\nonumber\\
p_y&=&\sin\gamma_2 \sin\theta_2 \cos\phi_2,
\nonumber\\
p_z&=&\sin\gamma_2 \sin\theta_2 \sin\phi_2,
\eeqa
where $0\le \phi_{1,2}\le 2\pi$, $0\le \theta_{1,2},\gamma_{1,2}\le \pi$, and all angles may be time dependent. The explicit expression of the Hamiltonian (\ref{Hr2}) in terms of these angles is in Appendix \ref{app2}.
We denote the initial and final states, at time $t=T$,  as $|\psi_r(0)\ra=a_1|1\ra+a_2|2\ra+a_3|3\ra+a_4|4\ra$ ($a_1^2+a_2^2+a_3^2+a_4^2=1$) and $|\psi_r(T)\ra=b_1|1\ra+b_2|2\ra+b_3|3\ra+b_4|4\ra$ ($b_1^2+b_2^2+b_3^2+b_4^2=1$), with phases $\varphi_k(0)=\epsilon_k$ and $\varphi_k(T)=\epsilon'_k$ ($k=2,3,4$).
Since $|\psi_r(T)\ra=U_r(T)|\psi_r(0)\ra$,  we have  four equations
\beq
\label{equations}
\begin{pmatrix}
b_1\\
b_2\\
b_3\\
b_4\\
\end{pmatrix}=U_r(T)
\begin{pmatrix}
a_1\\
a_2\\
a_3\\
a_4\\
\end{pmatrix}.
\eeq
If the angles at time $T$ and the initial $a_j$ components are fixed, these equations specify the final
coefficients $b_j$. Alternatively, if both initial and final coefficients are given, we have four equations and six variables to play with. The additional freedom may be used to cancel certain terms in the Hamiltonian
as demonstrated  below.
\subsection{The inverse tripod configuration}
As a first four-level system, we consider the ``Inverse Tripod'' configuration in Fig. \ref{config2}.
The three excited states ($|2\ra$, $|3\ra$ and $|4\ra$) are coupled to the ground state $|1\ra$ by
three couplings  $\Omega_{12}$, $\Omega_{13}$ and $\Omega_{14}$,
respectively \cite{tri1Bergmann,tri2Corbalan,tri3Knight}.
In this configuration, the transitions $|2\rangle \leftrightarrow |3\rangle$, $|2\rangle \leftrightarrow |4\rangle$ and $|3\rangle \leftrightarrow |4\rangle$, are not allowed so we want to cancel these couplings in the Hamiltonian (\ref{Hr2}).
One posible choice to set $\Omega_{23}(t)=\Omega_{34}(t)=\Omega_{24}(t)=0$ is
\beqa
\label{c_tripod}
\phi_2&=&\phi_1=\phi, \nonumber \\
\theta_2&=&\theta_1=\theta, \nonumber \\
\gamma_2&=&\gamma_1=\gamma(t),
\eeqa
see Eq. (\ref{Hrexp}), where $\phi$ and $\theta$ are constants and $\gamma(t)$ may generally depend on time.
The angles are equal for both isoclinic matrices, so
the evolution operator
\begin{widetext}
\beqa
\label{U_tripod}
U_r(t)&=&\cos{[2\gamma(t)]} |1\ra\la1|-2\cos{\gamma(t)}\sin{\gamma(t)} \cos{\theta}|1\ra\la2|-2\cos{\gamma(t)}\sin{\gamma(t)}\sin{\theta}\cos{\phi}|1\ra\la3|\nonumber\\
&-&2\cos{\gamma(t)} \sin{\gamma(t)} \sin{\theta} \sin{\phi} |1\ra\la4|
+\sin{[2\gamma(t)]} \cos{\theta}|2\ra\la1|+\{[\cos{\gamma(t)}]^2-\cos{(2\theta)}[\sin{\gamma(t)}]^2\}|2\ra\la2| \nonumber \\
&-&2[\sin{\gamma(t)}]^2\cos{\theta}\sin{\theta}\cos{\phi}|2\ra\la3|
-2[\sin{\gamma(t)}]^2\cos{\theta}\sin{\theta}\sin{\phi}|2\ra\la4|+\sin{[2\gamma(t)]} \sin{\theta} \cos{\phi}|3\ra\la1| \nonumber \\
&-&[\sin{\gamma(t)}]^2 \cos{\theta} \sin{\theta}\cos{\phi}|3\ra\la2|
+\{[\cos{\gamma(t)}]^2 +[\sin{\gamma(t)}]^2[(\cos{\theta})^2-\cos{(2\phi)}(\sin{\theta})^2]\}|3\ra\la3| \nonumber \\
&-&2[\sin{\gamma(t)}]^2(\sin{\theta})^2\cos{\phi}\sin{\phi}|3\ra\la4|
+ \sin{[2\gamma(t)]} \sin{\theta} \sin{\phi}|4\ra\la1|-2[\sin{\gamma(t)}]^2 \cos{\theta} \sin{\theta}\sin{\phi} |4\ra\la2| \nonumber \\
&-&2[\sin{\gamma(t)}]^2(\sin{\theta})^2\cos{\phi}\sin{\phi}|4\ra\la3|+\{[\cos{\gamma(t)}]^2+[\sin{\gamma(t)}]^2[(\cos{\theta})^2+\cos{(2\phi)}(\sin{\theta})^2] \}|4\ra\la4|
\eeqa
\end{widetext}
is a simple rotation (see a geometrical explanation in Appendix \ref{app1}),
and the rotation Hamiltonian reduces to
\beqa
H_r(t)=&-&2 i \hbar \{[\dot{\gamma}(t) \cos{\theta}]|1\ra \la 2|+[\dot{\gamma}(t) \cos{\phi} \sin{\theta}]|1\ra \la 3| \nonumber\\
&+&[\dot{\gamma}(t) \sin{\phi} \sin{\theta}]|1\ra \la 4|\}+H.c..
\eeqa
For this particular case, the couplings
\beqa
\Omega_{12}(t)&=& \dot \gamma(t)\cos{\theta}, \nonumber \\
\Omega_{13}(t)&=& \dot \gamma(t)\sin{\theta}\cos{\phi}, \nonumber \\
\Omega_{14}(t)&=&\dot \gamma(t)\sin{\theta}\sin{\phi},
\label{3omegas}
\eeqa
take the form of cartesian coordinates of a point on a sphere in terms of spherical coordinates.
Starting from the ground state $|1\ra$ we have freedom to achieve any final state. Setting $a_1=1$, $a_2=a_3=a_4=0$ and substituting Eq. (\ref{U_tripod}) in Eq. (\ref{equations}) we get
\beqa
b_1&=&\cos{[2\gamma(T)]},
\nonumber \\
b_2&=&\sin{[2\gamma(T)]} \cos{\theta},
\nonumber \\
b_3&=&\sin{[2\gamma(T)]} \sin{\theta} \cos{\phi},
\nonumber \\
b_4&=&\sin{[2\gamma(T)]} \sin{\theta} \sin{\phi},
\nonumber \\
\eeqa
which we rewrite as
\beqa
\label{s_tripod}
b_1&=&A, \nonumber \\
b_2&=&B C, \nonumber \\
b_3&=&B D E, \nonumber \\
b_4&=& B D F,
\eeqa
with
\beq
\label{cv_tripod}
\begin{array}{ll}
A=\cos{[2\gamma(T)]}, & B=\sin{[2\gamma(T)]},\\
C=\cos{\theta}, & D=\sin{\theta}, \\
E=\cos{\phi}, & F=\sin{\phi},
\end{array}
\eeq
obeying the conditions  $A^2+B^2=1$, $C^2+D^2=1$ and $E^2+F^2=1$.
The  system in Eq. (\ref{s_tripod}) with the above conditions has  solution
\beq
\label{sol_tripod}
\begin{array}{ll}
A=b_1, & B=\sqrt{b_2^2+b_3^2+b_4^2},
\\
C=\frac{b_2}{\sqrt{b_2^2+b_3^2+b_4^2}}, & D=\frac{\sqrt{b_3^2+b_4^2}}{\sqrt{b_2^2+b_3^2+b_4^2}},
\\
E=\frac{b_3}{\sqrt{b_3^2+b_4^2}}, & F=\frac{b_4}{\sqrt{b_3^2+b_4^2}},
\end{array}
\eeq
where we take positive square roots,
so it is possible to drive population transfers between the ground state and any final state.
To exemplify the method, let us implement the transition $|1\ra \rightarrow \frac{1}{\sqrt{3}}(|2\ra+|3\ra+|4\ra)$.
Substituting $b_1=0$, $b_2=1/\sqrt{3}$, $b_3=1/\sqrt{3}$ and $b_4=1/\sqrt{3}$ in Eq. (\ref{sol_tripod}) and using Eq. (\ref{cv_tripod})
we get four equations for $\gamma(T)$, $\theta$ and $\phi$ with solutions
\beq
\gamma(T)=\frac{\pi}{4}, \, \, \theta=\arctan \sqrt{2}, \, \, \phi=\frac{\pi}{4}.
\eeq
We now use an {ansatz} for
$\gamma(t)$ consistent with $\gamma(T)$, $\gamma(t)=\frac{\pi}{8}[1-\cos(\frac{\pi t}{T})]$.
It will determine the time-dependence of the Hamiltonian by Eq. (\ref{3omegas}).
Notice that this is just a simple choice, we could use different
functions, e.g. to optimize some physically relevant variable or improve robustness.

\begin{figure}[]
\scalebox{0.7}[0.7]{\includegraphics{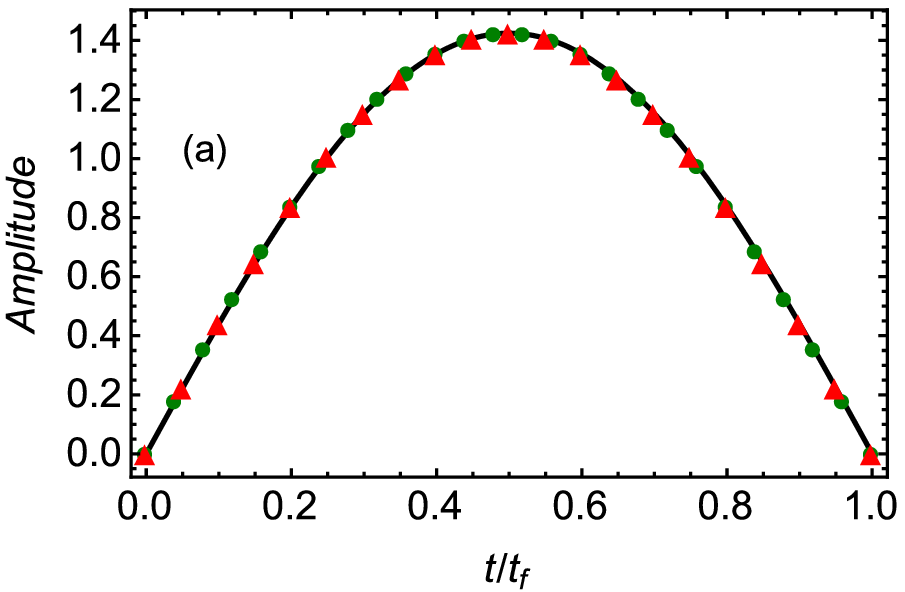}}
\scalebox{0.7}[0.7]{\includegraphics{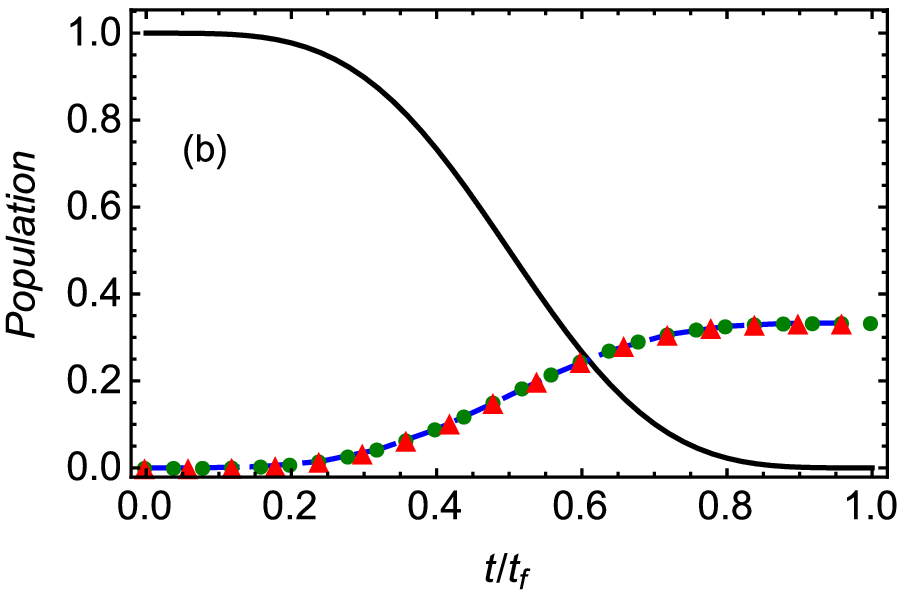}}
\caption{(Color online) (a) Overlapping couplings
$\Omega_{12}(t)$ (solid black line), $\Omega_{13}(t)$ (green dots) and $\Omega_{14}(t)$ (red triangles). (b) Populations of $|1\ra$ (solid black line),
$|2\ra$ (long-dashed blue line), $|3\ra$ (green dots) and $|4\ra$ (red triangles).
Parameters: $\phi=\frac{\pi}{4}$, $\theta=\arctan \sqrt{2}$, $\epsilon_k=0$ and $\epsilon_k'=\pi/3$, for $k=2,3,4$.
}
\label{fig2}
\end{figure}

For the phases we  use simple linear interpolation {ansatzes},
\beq
\label{phases_1}
\varphi_k(t)=\epsilon_k+\Delta_k t,
\eeq
where
\beq
\label{Deltak}
\Delta_k=(\epsilon_k'-\epsilon_k)/T,\; (k=2,3,4)
\eeq
may be interpreted as constant detunings in a quantum-optical realization, see Appendix C.
Substituting them in Eq. (\ref{H2}), the total Hamiltonian is
\beqa
\label{H3}
H(t)=&-&\hbar \left \{\sum_{k=2}^4 \Delta_k |k\ra \la k| + i[e^{-i (\epsilon_2+\Delta_2 t)}\Omega_{12}(t)|1\ra \la 2|\right.
\nonumber \\
&+&e^{-i (\epsilon_3+\Delta_3 t)} \Omega_{13}(t)|1\ra \la 3|
\nonumber\\
&+&e^{-i (\epsilon_4+\Delta_4 t)}\Omega_{14}(t)|1\ra \la 4|
\nonumber\\
&+&e^{i [(\epsilon_2-\epsilon_3)+(\Delta_2-\Delta_3)t]}\Omega_{23}(t)|2\ra \la 3|
\nonumber\\
&+&e^{i [(\epsilon_2-\epsilon_4)+(\Delta_2-\Delta_4)t]}\Omega_{24}(t)|2 \ra \la 4|
\nonumber\\
&+& \left. e^{i [(\epsilon_3-\epsilon_4)+(\Delta_3-\Delta_4)t]}\Omega_{34}(t)|3\ra \la 4|]\right \}+ H.c..
\nonumber \\
\eeqa
As an example, let us choose the following boundary conditions,
\beqa
\epsilon_k&=&0,
\nonumber \\
\epsilon_k'&=&\frac{\pi}{3},
\eeqa
$k=2,3,4$, to set the phases $\varphi_k(t)$.
Fig. \ref{fig2} (a) shows the common smooth amplitude of the couplings,
and Fig. \ref{fig2} (b) demonstrates the perfect population transfer.
%
%
%
\subsection{The diamond configuration}
%
%
%
\begin{figure}[]
\scalebox{0.5}[0.5]{\includegraphics{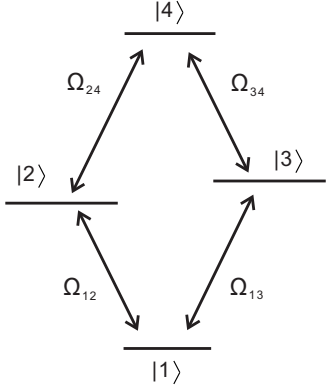}}
\caption{Energy level scheme for the diamond-type configuration with four couplings $\Omega_{12}$, $\Omega_{13}$, $\Omega_{24}$ and $\Omega_{34}$.}
\label{config3}
\end{figure}
\begin{figure}[]
\scalebox{0.7}[0.7]{\includegraphics{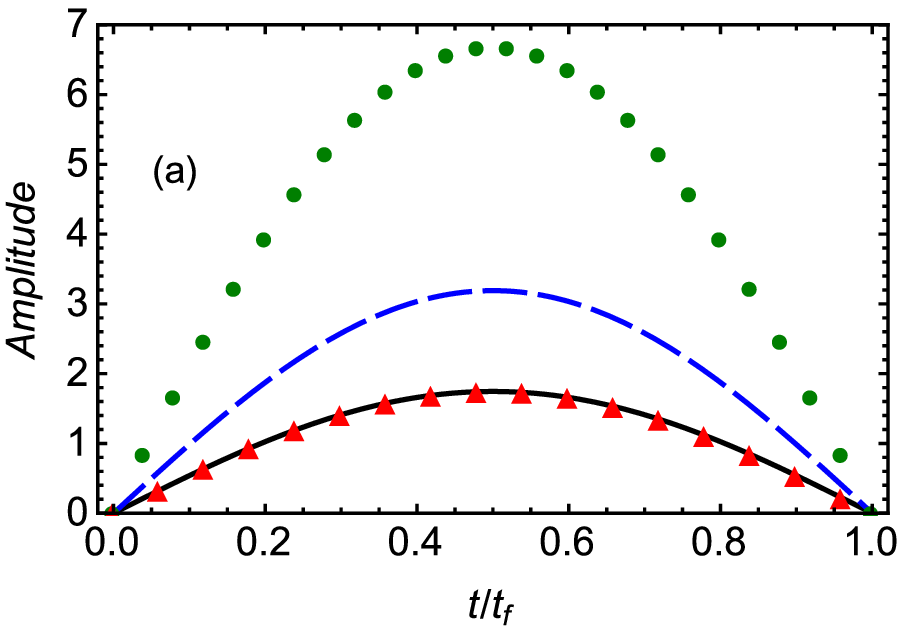}}
\scalebox{0.7}[0.7]{\includegraphics{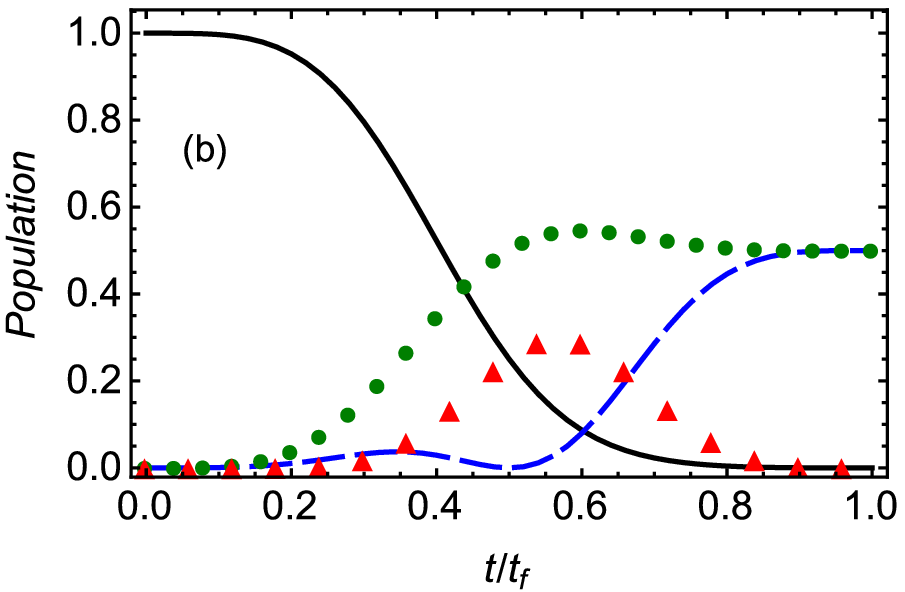}}
\caption{(Color online) (a) Couplings  $\Omega_{12}(t)$ (solid black line), $\Omega_{13}(t)$ (long-dashed blue line), $\Omega_{24}(t)$ (green dots) and $\Omega_{34}(t)$ (red triangles), $\Omega_{12}(t)=\Omega_{34}(t)$. (b) Populations of $|1\ra$ (solid black line), $|2\ra$ (long-dashed blue line), $|3\ra$ (green dots) and $|4\ra$ (red triangles).  The parameters are: $\phi_1=\phi_2=0$, $\theta_1=\frac{\pi}{2}$, $\theta_2=-\frac{3\pi}{4}$, $\epsilon_k=0$, $\epsilon_2'=\epsilon_4'=0$ and $\epsilon_3'=\pm \pi/2$. 
}
\label{fig3}
\end{figure}

%
Now we will focus on the diamond configuration shown in Fig. \ref{config3}.
In this configuration one ground state $|1\ra$ is coupled in a $V$-type structure to two intermediate states $|2\ra$, $|3\ra$, which are themselves coupled to a common excited state $|4\ra$ in a $\lambda$-type structure (see examples in atomic systems in Refs. \cite{D1Haim,D3Giovanna,D4Bao-Quan Ou} and in optical lattices in \cite{D2Thomas}).
Figure \ref{config3} shows that the transitions $|1\ra \leftrightarrow |4\ra$ and $|2\ra \leftrightarrow |3\ra$
are not allowed so, they must be cancelled in the Hamiltonian (\ref{Hr2}). To remove the unwanted terms we proceed similarly as in the inverse tripod, taking
now
\beqa
\label{c_diamond}
\phi_1&=&\phi_2=0, \nonumber \\
\dot{\theta}_1&=&\dot{\theta}_2=\dot{\phi}_1=\dot{\phi}_2=0,
\eeqa
to achieve $\Omega_{14}(t)=\Omega_{23}(t)=0$, which gives for the other couplings
\beqa
\label{couplings_diamond}
\Omega_{12}(t)&=&-[\dot{\gamma}_1(t) \cos\theta_1+\dot{\gamma}_2 (t) \cos\theta_2],
\nonumber\\
\Omega_{13}(t)&=&-[\dot{\gamma}_1(t) \sin\theta_1+\dot{\gamma}_2(t) \sin\theta_2],
\nonumber\\
\Omega_{24}(t)&=&\dot{\gamma}_1(t) \sin\theta_1-\dot{\gamma}_2(t) \sin\theta_2,
\nonumber\\
\Omega_{34}(t)&=&-[\dot{\gamma}_1(t) \cos\theta_1-\dot{\gamma}_2(t) \cos\theta_2].
\eeqa
The evolution operator becomes
\begin{widetext}
\begin{small}
\beqa
\label{U_diamond}
U_r(t)&=&[\cos{\gamma_1(t)}\cos{\gamma_2(t)}-\cos{(\theta_1-\theta_2)}\sin{\gamma_1(t)}\sin{\gamma_2(t)}]|1\ra\la1|
-[\sin{\gamma_1(t)}\cos{\gamma_2(t)}\cos{\theta_1}+\cos{\gamma_1(t)}\sin{\gamma_2(t)}\cos{\theta_2}]|1\ra\la2| \nonumber\\
&-&[\sin{\gamma_1(t)}\cos{\gamma_2(t)}\sin{\theta_1}+\cos{\gamma_1(t)}\sin{\gamma_2(t)}\sin{\theta_2}]|1\ra\la3|
-[\sin{\gamma_1(t)}\sin{\gamma_2(t)}\sin{(\theta_1-\theta_2)}]|1\ra\la4|\nonumber\\
&+&[\sin{\gamma_1(t)}\cos{\gamma_2(t)}\cos{\theta_1}+\cos{\gamma_1(t)}\sin{\gamma_2(t)}\cos{\theta_2}]|2\ra\la1|
+[\cos{\gamma_1(t)}\cos{\gamma_2(t)}-\cos{(\theta_1+\theta_2)}\sin{\gamma_1(t)}\sin{\gamma_2(t)}]|2\ra\la2|\nonumber\\
&-&[\sin{\gamma_1(t)}\sin{\gamma_2(t)}\sin{(\theta_1+\theta_2)}]|2\ra\la3|
+[\sin{\gamma_1(t)}\cos{\gamma_2(t)}\sin{\theta_1}-\cos{\gamma_1(t)}\sin{\gamma_2(t)}\sin{\theta_2}]|2\ra\la4| \nonumber\\
&+&[\sin{\gamma_1(t)}\cos{\gamma_2(t)}\sin{\theta_1}+\cos{\gamma_1(t)}\sin{\gamma_2(t)}\sin{\theta_2}]|3\ra\la1|
-[\sin{\gamma_1}\sin{\gamma_2}\sin{(\theta_1+\theta_2)}]|3\ra\la2|\nonumber\\
&+&[\cos{\gamma_1(t)}\cos{\gamma_2(t)}+\cos{(\theta_1+\theta_2)}\sin{\gamma_1(t)}\sin{\gamma_2(t)}]|3\ra\la3|
-[\sin{\gamma_1(t)}\cos{\gamma_2(t)}\cos{\theta_1}-\cos{\gamma_1(t)}\sin{\gamma_2(t)}\cos{\theta_2}]|3\ra\la4|\nonumber\\
&-&[\sin{\gamma_1(t)}\sin{\gamma_2(t)}\sin{(\theta_1-\theta_2)}]|4\ra\la1|
-[\sin{\gamma_1(t)}\cos{\gamma_2(t)}\sin{\theta_1}-\cos{\gamma_1(t)}\sin{\gamma_2(t)}\sin{\theta_2}]|4\ra\la2|\nonumber\\
&+&[\sin{\gamma_1(t)}\cos{\gamma_2(t)}\cos{\theta_1}-\cos{\gamma_1(t)}\sin{\gamma_2(t)}\cos{\theta_2}]|4\ra\la3|
+[\cos{\gamma_1(t)}\cos{\gamma_2(t)}+\cos{(\theta_1-\theta_2)}\sin{\gamma_1(t)}\sin{\gamma_2(t)}]|4\ra\la4|, \nonumber \\
\eeqa
\end{small}
\end{widetext}
and the rotating Hamiltonian is
\beqa
H_r(t)=&-& i \hbar \{[\dot\gamma_1(t)\cos{\theta_1}+\dot\gamma_2(t)\cos{\theta_2}]|1\ra \la 2|
\nonumber\\
&+&[\dot\gamma_1(t)\sin{\theta_1}+\dot\gamma_2(t)\sin{\theta_2}]|1\ra \la 3|
\nonumber\\
&+&[-\dot\gamma_1(t)\sin{\theta_1}+\dot\gamma_2(t)\sin{\theta_2}]|2\ra \la 4|
\nonumber\\
&+&[\dot\gamma_1(t)\cos{\theta_1}-\dot\gamma_2(t)\cos{\theta_2}]|3\ra \la 4|\}+H.c..
\nonumber\\
\eeqa
To design the Hamiltonian for a transition from $|\psi_r(0)\ra=|1\ra$, we set $a_1=1$, $a_2=a_3=a_4=0$ and substitute Eq. (\ref{U_diamond}) in Eq. (\ref{equations}),
\beqa
\label{system1_diamond}
b_1&=&\cos{\gamma_1(T)}\!\cos{\gamma_2(T)}-\cos{(\theta_1\!-\!\theta_2)}\!\sin{\gamma_1(T)}\!\sin{\gamma_2(T)},
\nonumber \\
b_2&=&\sin{\gamma_1(T)}\!\cos{\gamma_2(T)}\!\cos{\theta_1}+\cos{\gamma_1(T)}\!\sin{\gamma_2(T)}\!\cos{\theta_2},
\nonumber \\
b_3&=&\sin{\gamma_1(T)}\!\cos{\gamma_2(T)}\sin{\theta_1}+\cos{\gamma_1(T)}\!\sin{\gamma_2(T)}\!\sin{\theta_2},
\nonumber \\
b_4&=&-\sin{\gamma_1(T)}\sin{\gamma_2(T)}\sin{(\theta_1-\theta_2)}.
\nonumber \\
\eeqa
Using the  change of variables
\beq
\label{cv_diamond}
\begin{array}{ll}
A=\cos{\gamma_1(T)}, & B=\cos{\gamma_2(T)},\\
C=\sin{\gamma_1(T)}, & D=\sin{\gamma_2(T)}, \\
E=\cos{\theta_1}, & F=\cos{\theta_2}, \\
G=\sin{\theta_1}, & H=\sin{\theta_2},
\end{array}
\eeq
the equations in (\ref{system1_diamond}) become
\beqa
\label{s_diamond}
b_1&=&AB-CD(EF+GH), \nonumber \\
b_2&=&CBE+ADF, \nonumber \\
b_3&=&CBG+ADH, \nonumber \\
b_4&=&CD (HE-GF),
\eeqa
where $A^2+C^2=1$, $B^2+D^2=1$, $E^2+G^2=1$ and $F^2+H^2$.
The solution in terms of the final state coefficients is
\begin{widetext}
\beqa
\label{sol_diamond}
A&=&\frac{b_3 E-b_2G}{\sqrt{b_4^2+(b_3E-b_2G)^2}},
\nonumber \\
B&=&\frac{[(b_1 b_3+b_2 b_4)E+(b_3 b_4-b_1 b_2)G]\sqrt{b_4^2+(b_3E-b_2G)^2}}{(b_3^2+b_4^2)E^2-2 b_2 b_3 EG +(b_2^2+b_4^2)G^2},
\nonumber \\
C&=&\frac{b_4}{\sqrt{b_4^2+(b_3E-b_2G)^2}},
\nonumber \\
D&=&\sqrt{1-\frac{[(b_1 b_3+b_2 b_4)E+(b_3 b_4-b_1 b_2)G]^2[b_4^2+(b_3E-b_2G)^2]}{[(b_3^2+b_4^2)E^2-2 b_2 b_3 EG +(b_2^2+b_4^2)G^2]^2}},
\nonumber \\
F&=&-\frac{[(b_4 b_1-b_2 b_3)E+(b_2^2+b_4^2)G]\sqrt{b_4^2+(b_3E-b_2G)^2}}{[(b_3^2+b_4^2)E^2-2 b_2 b_3 EG +(b_2^2+b_4^2)G^2]D},
\nonumber \\
H&=&\frac{[(b_3^2+b_4^2)E-(b_2 b_3+b_1 b_4)G]\sqrt{b_4^2+(b_3E-b_2G)^2}}{[(b_3^2+b_4^2)E^2-2 b_2 b_3 EG +(b_2^2+b_4^2)G^2]D},
\eeqa
\end{widetext}
where $E$ and $G$ must obey $E^2+G^2=1$,
so there is freedom to fix the value of the angle $\theta_1$, see Eq. (\ref{cv_diamond}).
The other angles, $\gamma_{1,2}(T)$ and $\theta_2$,  are found from Eq. (\ref{cv_diamond}).
As an example, we study the population transfer from $|1\ra$ to the
final state $|\psi(T)\ra=\frac{1}{\sqrt{2}}(|2\ra\pm i|3\ra)$.
Substituting $b_1=0$, $b_2=1/\sqrt{2}$, $b_3=1/\sqrt{2}$ and $b_4=0$ in Eq. (\ref{sol_diamond}), choosing $\theta_1=\pi/2$ and using Eq. (\ref{cv_diamond}) we
find for the angles the values
\beq
\gamma_1(T)=\pi, \, \, \gamma_2(T)=\frac{\pi}{2}, \, \, \theta_2=-\frac{3\pi}{4}.
\eeq
For $\gamma_1(t)$ and $\gamma_2(t)$ we pick out smooth functions consistent with the  values
at $T$,
\beqa
\gamma_1(t)&=&\frac{\pi}{2}\left[1-\cos \left(\frac{\pi t}{T}\right)\right],
\nonumber \\
\gamma_2(t)&=&\frac{\pi}{4}\left[1-\cos \left(\frac{\pi t}{T}\right)\right].
\eeqa
To find  the full  Hamiltonian we use Eq. (\ref{H2})
with
$
\varphi_k(t)=\epsilon_k + \Delta_k t,\;\;k=2,3,4,
$
where the $\Delta_k$ are chosen to satisfy the boundary conditions of the example,
\beqa
\epsilon_k&=&0,
\nonumber \\
\epsilon_2'&=&0,\epsilon_3'=\pm \pi/2,\epsilon_4'=0.
\eeqa
The results are shown in Fig. (\ref{fig3}).
Figure \ref{fig3} (b) shows the perfect population transfer.
%
%

\subsection{The $\emph{N}$-type configuration}
%
\begin{figure}[]
\scalebox{0.6}[0.6]{\includegraphics{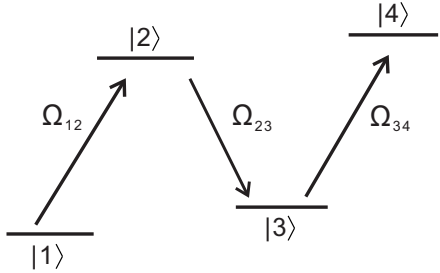}}
\caption{Energy level scheme for the four-level $\emph{N}$-configuration. There are three allowed couplings, $\Omega_{12}$, $\Omega_{23}$, and $\Omega_{34}$.}
\label{config1}
\end{figure}
%
\begin{figure}[hb]
\scalebox{0.7}[0.7]{\includegraphics{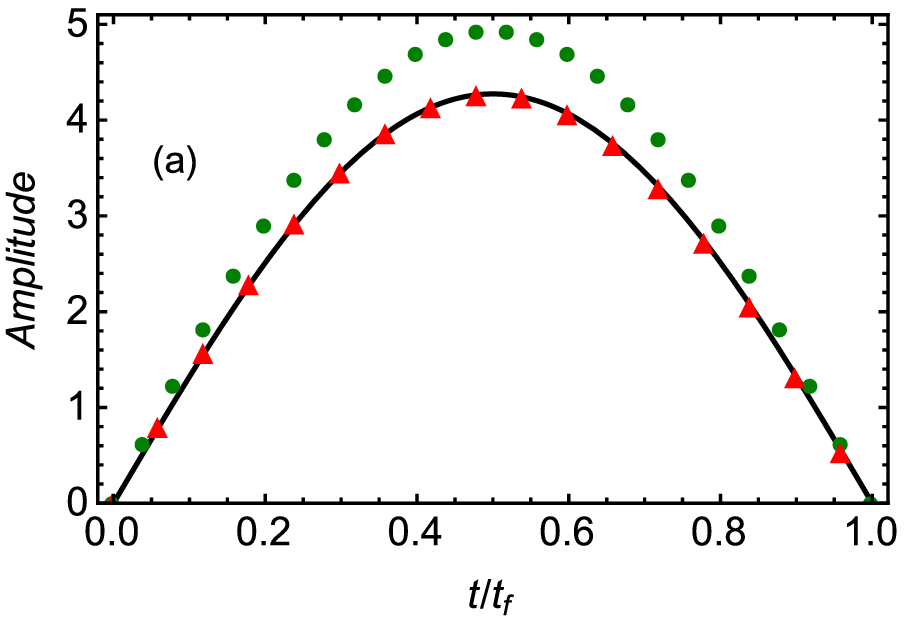}}
\scalebox{0.7}[0.7]{\includegraphics{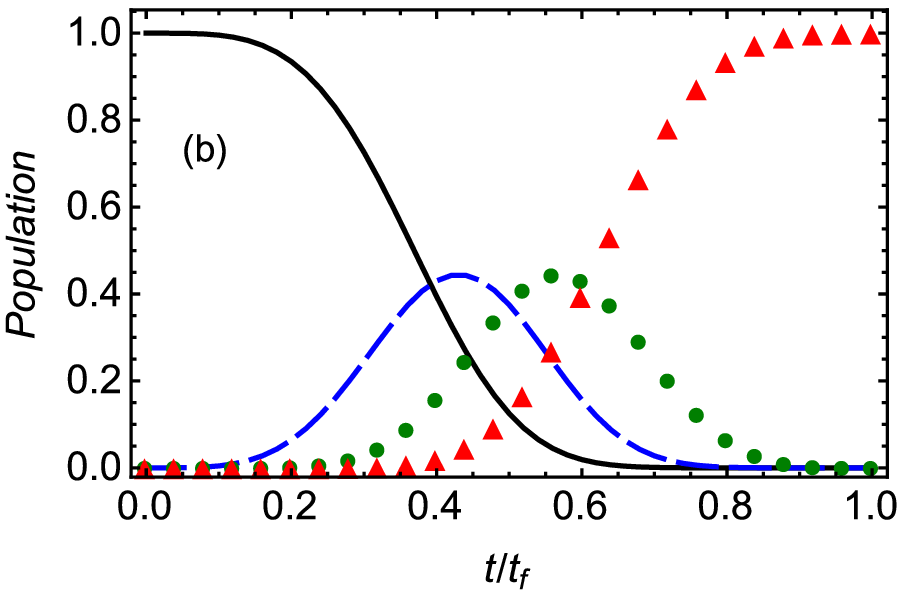}}
\caption{(Color online) (a) Couplings  $\Omega_{12}(t)$ (solid black line), $\Omega_{23}(t)$ (green dots),
 and $\Omega_{34}(t)$ (red triangles). (b) Populations  of $|1\ra$ (solid black line), $|2\ra$ (long-dashed blue line), $|3\ra$ (green dots) and $|4\ra$ (red triangles).
 The parameters are $\theta_1=\pi/6$, $\theta_2=\pi/2$, $\gamma_1(T)=\pi$, $\gamma_2(T)=-\pi/2$, $\epsilon_2=0$, $\epsilon_3=0$, $\epsilon_4=0$, $\epsilon_2'=0$, $\epsilon_3'=0$, $\epsilon_4'=\pi/6$.
}
\label{fig1}
\end{figure}
%
The last four-level structure  we  study is the $\emph{N}$-type level scheme \cite{N1Xu}, with three non-zero couplings $\Omega_{12}$,
$\Omega_{23}$ and $\Omega_{34}$, see Fig. \ref{config1} (A ladder configuration would be treated similarly.). This configuration
is applied, for example, to realize the phenomenon of EIT and population transfers in optical lattice systems
\cite{N1Xu,N2Goren,N3zhu}.  To eliminate the unwanted terms, i.e.,
to have $\Omega_{13}(t)=\Omega_{14}(t)=\Omega_{24}(t)=0$ in Eq. (\ref{Hr2})
one possible solution is
\beqa
\label{s_Ntype}
\dot{\phi}_1&=&\dot{\phi}_2=\dot{\theta}_1=\dot{\theta}_2=0,
\\
\phi_1&=&\phi_2=\frac{\pi}{2},
\\
\dot{\gamma}_1&=&-\frac{\sin\theta_2}{\sin\theta_1}\dot{\gamma}_2.
\label{gaga}
\eeqa
The Hamiltonian $H_r(t)$ becomes
\beqa
H_r(t)&=&i[(\cot\theta_1 \sin\theta_2-\cos\theta_2)\dot{\gamma}_2(t)|1\ra \la2| \nonumber
\\
&+&2\sin\theta_2 \dot{\gamma}_2(t) |2\ra \la3| \nonumber \\
&+&(\cot\theta_1 \sin\theta_2+\cos\theta_2)\dot{\gamma}_2(t) |3\ra \la4|]+ H.c. \nonumber
\\
\eeqa
and the couplings are
\beqa
\label{couplings_Ntype}
\Omega_{12}(t)&=&\dot\gamma_2(t) (\sin{\theta_2} \cot{\theta_1}-\cos{\theta_2}), \nonumber \\
\Omega_{23}(t)&=&2 \dot \gamma_2(t) \sin{\theta_2}, \nonumber \\
\Omega_{34}(t)&=&\dot\gamma_2(t) (\sin{\theta_2} \cot{\theta_1}+\cos{\theta_2}).
\eeqa
Unlike the previous cases, we do not find an analytical expression for the general solution of
$U_r(T)$ in Eq. (\ref{equations}) for  the initial state $|\psi_r(0)\ra=|1\ra$. However, for a given final state the system can be  solved
to get the needed angles.
As an example, let us engineer the interaction to go from $|\psi_r(0)\ra=|1\ra$ to  $|\psi_r(T)\ra=|4\ra$. From  Eq. (\ref{equations}) and Eq. (\ref{Urexp}), we get four equations
for $\gamma_1(T)$, $\gamma_2(T)$, [note that $\gamma_1=-\frac{\sin\theta_2}{\sin{\theta_1}}\gamma_2+c$, see Eq. (\ref{gaga})], $\theta_1$, and $\theta_2$ with solutions
$\theta_1=\pi/6$, $\theta_2=\pi/2$, $\gamma_2(T)=-\pi/2$, $\gamma_1(T)=\pi$.
We choose again $\gamma_2(t)=\frac{\pi}{4}[\cos(\frac{\pi t}{T})-1]$ as a smooth {ansatz},  so $H_r(t)$ takes the form
\beq
H_r(t)=i(\Omega_{12}(t)|1\ra \la 2|+\Omega_{23}(t)|2\ra \la 3|+\Omega_{34}(t)|3\ra \la 4|)+H.c.,
\eeq
where
\beqa
\Omega_{12}(t)&=&\Omega_{34}(t)=-\frac{\sqrt{3} \pi^2}{4 T} \sin\left(\frac{\pi t}{T}\right),
\nonumber\\
\Omega_{23}(t)&=&-\frac{\pi^2}{2 T} \sin\left(\frac{\pi t}{T}\right).
\eeqa
We may use the simple linear interpolation (\ref{phases_1}) for the phases.  For an example with  boundary conditions
\beqa
\epsilon_k&=&0, \nonumber
\\
\epsilon_2'&=&\epsilon_3'=0,\;\epsilon_4'=\pi/6,
\eeqa
Fig. \ref{fig1} shows the couplings (a) and population transfer (b) from state  $|1\ra$ to the desired state $e^{i \epsilon_4'}|4\ra$.
\section{Discussion}
We have set a method to design four-level Hamiltonians so as to drive, in principle in an arbitrary time,  specific transitions
for different, preselected configurations of the couplings. For arbitrary final states, the method requires full control of the
real and imaginary parts of the couplings, and of constant energy shifts. The possibility to realize this level of control
will depend on the specific system and physical realization of the
Hamiltonian (\ref{H2}). In an atomic system subjected to optical laser fields, this is an interaction picture Hamiltonian
after applying the rotating wave approximation, see Appendix C, where the diagonal terms can be interpreted as detunings,
and the non-diagonal terms as complex Rabi frequencies. Independent control may be required of the real and imaginary
parts of the Rabi frequencies for  final states with  non-zero phases.

We intend to apply these results in different scenarios. 
For example, to manipulate the spin state in quantum dots with spin-orbit coupling and electric field control \cite{Ban2012}.  
As for generalizations, the geometry of rotations in higher dimensions has been much less studied that in 3D or 4D, but
there are different approaches available \cite{nD1,nD2} that could be used to generalize the current scheme to
systems with more levels.

\acknowledgments{We are grateful to E. Y. Sherman and D. Gu\'ery-Odelin for useful comments on the manuscript. 
This work was supported by the Basque Country Government (Grant No. IT986-16),  and MINECO/FEDER,UE (Grant No. FIS2015-67161- P).}

\appendix
\section{Quaternions and 4D rotations}
\label{app1}
A quaternion \textbf{q}  can be defined as the sum of a scalar $q_w$ and a vector $\vec{q}$,
namely \cite{qua1Wilkens}
\beq
\textbf{q}=q_w+\vec{q}=q_w+q_x \textbf{\textit{i}}+q_y \textbf{\textit{j}}+q_z \textbf{\textit{k}}.
\eeq
The rule of product of two quaternions is defined by
\beq
\begin{split}
\textbf{\textit{i}}^2=\textbf{\textit{j}}^2&=\textbf{\textit{k}}^2=\textbf{\textit{i}}\textbf{\textit{j}}\textbf{\textit{k}}=-1.
\end{split}
\eeq
If  $|\textbf{q}|^2=1$, namely, $q_w^2+q_x^2+q_y^2+q_z^2=1$,  $\textbf{q}$ is a unit quaternion and $\textbf{q}^{-1}=\bar{q}$. If $\textbf{u}=\vec{u}$ and $|\textbf{u}|^2=1$, $\textbf{u}$ is a pure unit quaternion, and every pure unit quaternion is a square root of -1.
A unit quaternion can be expressed
in terms of a real number $\gamma$ and a pure unit quaternion $\textbf{u}$ as
\beq
\textbf{q}=e^{\textbf{u} \gamma}=\cos\gamma+\textbf{u} \sin \gamma.
\eeq
Consider two arbitrary unit quaternions $\textbf{p}$ and $\textbf{q}$. We may choose proper pure unit quaternions $\textbf{u}$ and $\textbf{v}$ with corresponding real numbers $\gamma_1$ and $\gamma_2$, so that $\textbf{p}=e^{\textbf{u} \gamma_1}$ and $\textbf{q}=e^{\textbf{v} \gamma_2}$. As  noted in Sec. \ref{subsectionA},
an arbitrary rotation $R$  in $\mathbb{E}^4$ of a 4-vector $C$ can be represented by the product $qCp$, associated with left and right isoclinic rotations with rotation angles $\gamma_1$ and $\gamma_2$. $R$ also corresponds to a product of rotations in two mutually orthogonal planes \cite{iso1Federico,iso2Thomas,qua2Mebius1,qua3Mebius2}. If $\textbf{u}\neq\pm\textbf{v}$, $R$ rotates the plane spanned by $\textbf{u}+\textbf{v}$ and $\textbf{u}\textbf{v}-1$ through the angle $|\gamma_1+\gamma_2|$, and the plane spanned by $\textbf{v}-\textbf{u}$ and $\textbf{u}\textbf{v}+1$ through the angle $|\gamma_1-\gamma_2|$, respectively \cite{qua1Wilkens}.
If $\textbf{u}=\pm\textbf{v}$, the planes are spanned by $1$ and $\textbf{u}$ and its orthogonal complement, and the rotation angles are  as well $|\gamma_1+\gamma_2|$ and $|\gamma_1-\gamma_2|$ \cite{qua1Wilkens}.
\section{Hamiltonian and evolution}
\label{app2}
Using Eqs. (\ref{rotationmatrix},\ref{Hr2},\ref{parametrization}), the parameterized Hamiltonian is given by
\begin{widetext}
\begin{small}
\beqa
\label{Hrexp}
&&H_r(t)=i \hbar \dot{U}_r(t) U^{\dagger}_r(t)
\nonumber\\
&&=i\hbar\left[\sin\gamma_1 \sin\theta_1 (\dot{\theta}_1 \cos\gamma_1 - \dot{\phi}_1 \sin\gamma_1\sin\theta_1 )+\sin\gamma_2 \sin\theta_2 (\dot{\theta}_2 \cos\gamma_2 + \dot{\phi}_2  \sin\gamma_2 \sin\theta_2)-\dot{\gamma}_1\cos\theta_1 - \dot{\gamma}_2\cos\theta_2 \right]|1\ra \la 2|
\nonumber\\
&&+i\hbar\left[\dot{\theta}_1 \sin\gamma_1( \sin\gamma_1\sin\phi_1 -\cos\gamma_1 \cos\theta_1 \cos\phi_1)
-\dot{\theta}_2 \sin\gamma_2(\sin\gamma_2\sin\phi_2  +\cos\gamma_2 \cos\theta_2 \cos\phi_2 ) - \dot{\gamma}_1 \sin\theta_1\cos\phi_1 \right.
\nonumber\\
&&\left. -\dot{\gamma}_2\sin\theta_2\cos\phi_2+\dot{\phi}_1\sin\gamma_1 \sin\theta_1 (\cos\gamma_1 \sin\phi_1+\sin\gamma_1 \cos\theta_1 \cos\phi_1)+\dot{\phi}_2 \sin\gamma_2 \sin\theta_2  (\cos\gamma_2 \sin\phi_2 - \sin\gamma_2 \cos\theta_2 \cos\phi_2)\right]|1\ra \la 3|
\nonumber\\
&&+i\hbar\left[-\dot{\theta}_1 \sin\gamma_1( \sin\gamma_1\cos\phi_1 +\cos\gamma_1 \cos\theta_1 \sin\phi_1)
+\dot{\theta}_2 \sin\gamma_2(\sin\gamma_2\cos\phi_2  -\cos\gamma_2 \cos\theta_2 \sin\phi_2 ) - \dot{\gamma}_1 \sin\theta_1\sin\phi_1 \right.
\nonumber\\
&&\left. -\dot{\gamma}_2\sin\theta_2\sin\phi_2-\dot{\phi}_1\sin\gamma_1 \sin\theta_1 (\cos\gamma_1 \cos\phi_1-\sin\gamma_1 \cos\theta_1 \sin\phi_1)-\dot{\phi}_2 \sin\gamma_2 \sin\theta_2  (\cos\gamma_2 \cos\phi_2 + \sin\gamma_2 \cos\theta_2 \sin\phi_2)\right]|1\ra \la 4|
\nonumber\\
&&+i\hbar\left[-\dot{\theta}_1 \sin\gamma_1( \sin\gamma_1\cos\phi_1 +\cos\gamma_1 \cos\theta_1 \sin\phi_1)
-\dot{\theta}_2 \sin\gamma_2(\sin\gamma_2\cos\phi_2  -\cos\gamma_2 \cos\theta_2 \sin\phi_2 ) - \dot{\gamma}_1 \sin\theta_1\sin\phi_1 \right.
\nonumber\\
&&\left. +\dot{\gamma}_2\sin\theta_2\sin\phi_2-\dot{\phi}_1\sin\gamma_1 \sin\theta_1 (\cos\gamma_1 \cos\phi_1-\sin\gamma_1 \cos\theta_1 \sin\phi_1)+\dot{\phi}_2 \sin\gamma_2 \sin\theta_2  (\cos\gamma_2 \cos\phi_2 + \sin\gamma_2 \cos\theta_2 \sin\phi_2)\right]|2\ra \la 3|
\nonumber\\
&&+i\hbar\left[-\dot{\theta}_1 \sin\gamma_1( \sin\gamma_1\sin\phi_1 -\cos\gamma_1 \cos\theta_1 \cos\phi_1)
-\dot{\theta}_2 \sin\gamma_2(\sin\gamma_2\sin\phi_2  +\cos\gamma_2 \cos\theta_2 \cos\phi_2 ) +\dot{\gamma}_1 \sin\theta_1\cos\phi_1 \right.
\nonumber\\
&&\left. -\dot{\gamma}_2\sin\theta_2\cos\phi_2-\dot{\phi}_1\sin\gamma_1 \sin\theta_1 (\cos\gamma_1 \sin\phi_1+\sin\gamma_1 \cos\theta_1 \cos\phi_1)+\dot{\phi}_2 \sin\gamma_2 \sin\theta_2  (\cos\gamma_2 \sin\phi_2 - \sin\gamma_2 \cos\theta_2 \cos\phi_2)\right]|2\ra \la 4|
\nonumber\\
&&=i\hbar\left[\sin\gamma_1 \sin\theta_1 (\dot{\theta}_1 \cos\gamma_1 - \dot{\phi}_1 \sin\gamma_1\sin\theta_1 )-\sin\gamma_2 \sin\theta_2 (\dot{\theta}_2 \cos\gamma_2 + \dot{\phi}_2  \sin\gamma_2 \sin\theta_2)-\dot{\gamma}_1\cos\theta_1 + \dot{\gamma}_2\cos\theta_2 \right]|3\ra \la 4|
\nonumber\\
\eeqa
\end{small}
The time-dependent evolution operator parameterized by the generalized spherical angles in Eq. (\ref{parametrization}) is
\beq
\label{Urexp}
\begin{split}
U_r(t)=
&\{ \cos\gamma_1 \cos\gamma_2 - \sin\gamma_1 \sin\gamma_2 [\sin\theta_1 \sin\theta_2 \cos(\phi_1- \phi_2) + \cos\theta_1 \cos\theta_2] \} |1\ra \la 1|,\\
&\{ \sin\gamma_2 [\cos\theta_2 \cos\gamma_1 -\sin\theta_1 \sin\theta_2 \sin\gamma_1 \sin(\phi_1- \phi_2)] + \cos\theta_1 \sin\gamma_1\cos\gamma_2\}|2\ra \la 1|,\\
&\{ \sin\gamma_2 [\sin\gamma_1 (\sin\theta_1 \cos\theta_2 \sin\phi_1 -\cos\theta_1 \sin\theta_2 \sin\phi_2)+\sin \theta_2 \cos \gamma_1 \cos \phi_2]+\sin \theta_1 \sin \gamma_1 \cos \gamma_2 \cos \phi_1\}|3\ra \la 1|\\
&\{ \sin \gamma_2 [\sin \gamma_1 (\cos \theta_1 \sin \theta_2 \cos \phi_2-\sin \theta_1 \cos \theta_2 \cos \phi_1)+\sin \theta_2 \cos \gamma_1 \sin \phi_2]+\sin \theta_1 \sin \gamma_1 \cos \gamma_2 \sin \phi_1\}|4\ra \la 1|\\
&\{-\sin \gamma_2 [\sin \theta_1 \sin \theta_2 \sin \gamma_1 \sin ( \phi_1- \phi_2)+\cos \theta_2 \cos \gamma_1]-\cos \theta_1 \sin \gamma_1 \cos \gamma_2\}|1\ra \la 2|\\
&\{ \sin \gamma_1 \sin \gamma_2 [\sin \theta_1 \sin \theta_2 \cos ( \phi_1- \phi_2)-\cos \theta_1 \cos \theta_2]+\cos \gamma_1 \cos \gamma_2\}|2\ra \la 2|\\
&\{ \sin \theta_1 \sin \gamma_1 \cos \gamma_2 \sin \phi_1-\sin \gamma_2 [\sin \gamma_1 (\sin \theta_1 \cos \theta_2 \cos \phi_1+\cos \theta_1 \sin \theta_2 \cos \phi_2)+\sin \theta_2 \cos \gamma_1 \sin \phi_2]\}|3\ra \la 2|\\
&\{ \sin \theta_2 \cos \gamma_1 \sin \gamma_2 \cos \phi_2-\sin \gamma_1 [\sin \gamma_2 (\sin \theta_1 \cos \theta_2 \sin \phi_1+\cos \theta_1 \sin \theta_2 \sin \phi_2)+\sin \theta_1 \cos \gamma_2 \cos \phi_1]\}|4\ra \la 2|\\
&\{ \sin \theta_1 \sin \gamma_1 \cos \gamma_2 (-\cos \phi_1)-\sin \gamma_2 [\sin \gamma_1 (\cos \theta_1 \sin \theta_2 \sin \phi_2-\sin \theta_1 \cos \theta_2 \sin \phi_1)+\sin \theta_2 \cos \gamma_1 \cos \phi_2]\}|1\ra \la 3|\\
&\{ \sin \theta_2 \cos \gamma_1 \sin \gamma_2 \sin \phi_2- \sin \gamma_1 [\sin \gamma_2 (\sin \theta_1 \cos \theta_2 \cos \phi_1+\cos \theta_1 \sin \theta_2 \cos \phi_2)+\sin \theta_1 \cos \gamma_2 \sin \phi_1]\}|2\ra \la 3|\\
&\{ \sin \gamma_1 \sin \gamma_2 [\cos \theta_1 \cos \theta_2-\sin \theta_1 \sin \theta_2 \cos ( \phi_1+ \phi_2)]+\cos \gamma_1 \cos \gamma_2\}|3\ra \la 3|\\
&\{ \cos \theta_1 \sin \gamma_1 \cos \gamma_2-\sin \gamma_2 [\sin \theta_1 \sin \theta_2 \sin \gamma_1 \sin ( \phi_1+ \phi_2)+\cos \theta_2 \cos \gamma_1]\}|4\ra \la 3|\\
&\{ \sin \theta_1 \sin \gamma_1 (-\cos \gamma_2) \sin \phi_1-\sin \gamma_2 [\sin \gamma_1 (\sin \theta_1 \cos \theta_2 \cos \phi_1-\cos \theta_1 \sin \theta_2 \cos \phi_2)+\sin \theta_2 \cos \gamma_1 \sin \phi_2]\}|1\ra \la 4|\\
&\{ \sin \theta_1 \sin \gamma_1 \cos \gamma_2 \cos \phi_1-\sin \gamma_2 [\sin \gamma_1 (\sin \theta_1 \cos \theta_2 \sin \phi_1+\cos \theta_1 \sin \theta_2 \sin \phi_2)+\sin \theta_2 \cos \gamma_1 \cos \phi_2]\}|2\ra \la 4|\\
&\{ \cos \theta_2 \cos \gamma_1 \sin \gamma_2-\sin \gamma_1 [\sin \theta_1 \sin \theta_2 \sin \gamma_2 \sin ( \phi_1+ \phi_2)+\cos \theta_1 \cos \gamma_2]\}|3\ra \la 4|\\
&\{ \sin \gamma_1 \sin \gamma_2 [\sin \theta_1 \sin \theta_2 \cos ( \phi_1+ \phi_2)+\cos \theta_1 \cos \theta_2]+\cos \gamma_1 \cos \gamma_2\}|4\ra \la 4|.
\end{split}
\eeq
\end{widetext}
\section{Connection with quantum optics (diamond configuration)}
\label{q_o}
To relate  the Hamiltonian  of the inverse engineering approach, Eq.  (\ref{H2}),
to an interaction picture Hamiltonian for a four-level atom illuminated by laser fields,
we assume a semiclassical description of the interaction of the atom with coupling laser fields.
Neglecting atomic motion, the Hamiltonian in the Schr\"odinger picture for the diamond configuration and fields
composed by combinations of out-of-phase quadrature components is
\begin{widetext}
\beqa
H(t)&=& \hbar\left \{  \tilde\Omega_{12}(t) \left[  |1\ra \la2| + |2\ra \la1| \right] \cos{(\omega_{12} t+\phi_{12})}  -  \tilde\Omega'_{12}(t) \left[ |1\ra \la2| + |2\ra \la1| \right] \sin{(\omega_{12} t+\phi_{12})}   \right. \nonumber \\
&+&\tilde\Omega_{13}(t) \left[  |1\ra \la3| +  |3\ra \la1| \right] \cos{(\omega_{13} t+\phi_{13})}  -  \tilde\Omega'_{13}(t) \left[  |1\ra \la3| + |3\ra \la1| \right] \sin{(\omega_{13} t+\phi_{13})} \nonumber \\
&+&\tilde\Omega_{24}(t) \left[  |2\ra \la4| +   |4\ra \la2| \right] \cos{(\omega_{24} t+\phi_{24})}  -  \tilde\Omega'_{24}(t) \left[ |2\ra \la4| + |4\ra \la2| \right] \sin{(\omega_{24} t+\phi_{24})} \nonumber \\
&+&\tilde\Omega_{34}(t) \left[  |3\ra \la4| + |4\ra \la3| \right] \cos{(\omega_{34} t+\phi_{34})}  -  \tilde\Omega'_{34}(t) \left[  |3\ra \la4| +  |4\ra \la3| \right] \sin{(\omega_{34} t+\phi_{34})} \nonumber \\
&+& \sum_{i=2}^4  \omega_i |i\rangle \langle i|  \left. \right \},
\eeqa
\end{widetext}
where we use the vector basis $|1\rangle=\left ( \scriptsize{\begin{array} {rcccl} 1\\ 0\\0\\0 \end{array}} \right), |2\rangle=\left ( \scriptsize{\begin{array} {rcccl} 0\\ 2\\0\\0 \end{array}} \right),
|3\rangle=\left ( \scriptsize{\begin{array} {rcccl} 0\\ 0\\1\\0 \end{array}} \right), |4\rangle=\left ( \scriptsize{\begin{array} {rcccl} 0\\ 0\\0\\1 \end{array}} \right)$. $\tilde\Omega_{ij}(t),\tilde\Omega'_{ij}(t)$ are the  atom-field coupling strengths (Rabi frequencies), assumed real for simplicity,
and $\phi_{ij}$ the phases of the coherent driving fields. The atomic levels $|i\ra$ have energies $\hbar \omega_i$ and the fields have angular frequencies $\omega_{ij}$.
We choose the energy zero to match that of level $|1\ra$ ($\omega_1=0$).

To transform the system into a laser-adapted interaction picture (rotating frame), we define the unitary operator
\begin{widetext}
\beq
U_0(t)=
\begin{pmatrix}
1&0&0&0\\
0&e^{i(\omega_{12} t+\phi_{12})}&0&0\\
0&0&e^{i(\omega_{13} t+\phi_{13})}&0\\
0&0&0&e^{i[(\omega_{12}+\omega_{24}) t+\phi_{12}+\phi_{24}]}
\end{pmatrix}.
\eeq
\end{widetext}
Using
\beq
H_I(t)=U_0(t) H(t) U_0^\dag (t) + i \hbar \dot U_0(t) U_0^\dag(t),
\eeq
and imposing the four-photon resonance condition \cite{Buckle1986,Morigi2002,Arimondo2016}
\beq
\label{resonance_c}
\omega_{13}+\omega_{34}=\omega_{12}+\omega_{24},
\eeq
the Hamiltonian in the interacting picture is
\begin{widetext}
\beqa
H_I(t)&=&\frac{\hbar}{2} \left \{ 2(\omega_2-\omega_{12}) |2\ra \la 2|+2 (\omega_3-\omega_{13})|3\ra \la 3|+2 (\omega_4-\omega_{12}-\omega_{24})|4\ra \la 4| \right. \nonumber \\
&+&\tilde\Omega_{12}(t)\left [(1+e^{-2i(\omega_{12} t+\phi_{12})}) |1\ra \la2| +(1+e^{2i(\omega_{12} t+\phi_{12})}) |2\ra \la1| \right] \nonumber \\
&+&i \tilde\Omega'_{12}(t)\left [(1-e^{-2i(\omega_{12} t+\phi_{12})}) |1\ra \la2| - (1-e^{2i(\omega_{12} t+\phi_{12})}) |2\ra \la1| \right] \nonumber \\
&+&\tilde\Omega_{13}(t)\left [(1+e^{-2i(\omega_{13} t+\phi_{13})}) |1\ra \la3| +(1+e^{2i(\omega_{13} t+\phi_{13})}) |3\ra \la1| \right] \nonumber \\
&+&i \tilde\Omega'_{13}(t)\left [(1-e^{-2i(\omega_{13} t+\phi_{13})}) |1\ra \la3| - (1-e^{2i(\omega_{13} t+\phi_{13})}) |3\ra \la1| \right] \nonumber \\
&+&\tilde\Omega_{24}(t)\left [(1+e^{-2i(\omega_{24} t+\phi_{24})}) |2\ra \la4| +(1+e^{2i(\omega_{24} t+\phi_{24})}) |4\ra \la2| \right] \nonumber \\
&+&i \tilde\Omega'_{24}(t)\left [(1-e^{-2i(\omega_{24} t+\phi_{24})}) |2\ra \la4| -(1-e^{2i(\omega_{24} t+\phi_{24})}) |4\ra \la2| \right] \nonumber \\
&+& \tilde\Omega_{34}(t)\left[(1+e^{-2i(\omega_{34} t+\phi_{34})}) e^{-i \Phi} |3\ra \la4| + (1+e^{2i(\omega_{34} t+\phi_{34})}) e^{i \Phi} |4\ra \la3| \right] \nonumber \\
&+&\left. i \tilde\Omega'_{34}(t)\left [(1-e^{-2i(\omega_{34} t+\phi_{34})}) e^{-i \Phi} |3\ra \la4| - (1-e^{2i(\omega_{34} t+\phi_{34})})e^{i \Phi} |4\ra \la3| \right]  \right \}
\eeqa
\end{widetext}
where
\beq
\Phi=\phi_{12}-\phi_{13}+\phi_{24}-\phi_{34}.
\eeq
Applying now a rotating wave approximation (RWA) to get rid of the counter-rotating terms we end up with
\begin{widetext}
\begin{small}
\beq
\label{HIRWA}
H_{I,RWA}(t)=\frac{\hbar}{2}
\begin{pmatrix}
0&\tilde\Omega_{12}(t)+i\tilde \Omega'_{12}(t)&\tilde\Omega_{13}(t)+i\tilde \Omega'_{13}(t)&0\\
\tilde\Omega_{12}(t)-i\tilde \Omega'_{12}(t)&\tilde\Delta_2&0&\tilde\Omega_{24}(t)+i\tilde \Omega'_{24}(t)\\
\tilde\Omega_{13}(t)-i\tilde \Omega'_{13}(t)&0& \tilde\Delta_3&(\tilde\Omega_{34}(t)+i\tilde \Omega'_{34}(t)) e^{-i\Phi}\\
0&\tilde\Omega_{24}(t)-i\tilde \Omega'_{24}(t)&(\tilde\Omega_{34}(t)-i\tilde \Omega'_{34}(t)) e^{i\Phi}&\tilde \Delta_4
\end{pmatrix},
\eeq
\end{small}
\end{widetext}
where $\tilde\Delta_i$ ($i=2,3,4$) are the detunings defined as
\beqa
\label{detunings_qo}
\tilde\Delta_2&=&2(\omega_2-\omega_{12}), \nonumber \\
\tilde\Delta_3&=&2(\omega_3-\omega_{13}), \nonumber \\
\tilde\Delta_4&=&2(\omega_4-\omega_{12}-\omega_{24}).
\eeqa
Assuming that the phases of the coherent driving fields can be manipulated to satisfy
\beq
\phi_{12}-\phi_{13}+\phi_{24}-\phi_{34}=0,
\eeq
the Hamiltonian in Eq. (\ref{HIRWA}) has the structure of the one in Eq. (\ref{H2}).

Notice that, the four-photon resonance condition (\ref{resonance_c}) is key to find a simple Hamiltonian
structure in terms of the Rabi frequencies for closed-loop configurations.
Equating the diagonal terms, $-\Delta_i=\tilde\Delta_i/2$, the laser (angular) frequencies are
\beqa
\omega_{12}&=&\omega_2-\frac{\epsilon'_2-\epsilon_2}{2 T},
\nonumber \\
\omega_{13}&=&\omega_3-\frac{\epsilon'_3-\epsilon_3}{2 T},
\nonumber \\
\omega_{24}&=&\omega_4-\omega_2+\frac{\epsilon'_2-\epsilon_2}{2 T}-\frac{\epsilon'_4-\epsilon_4}{2 T},
\eeqa
and, to satisfy the four-photon resonance condition,
\beq
\omega_{34}=\omega_4-\omega_3-\frac{\epsilon'_4-\epsilon_4}{2 T}+\frac{\epsilon'_3-\epsilon_3}{2 T}.
\eeq
Comparing the non-diagonal terms in Eqs. (\ref{HIRWA}) and (\ref{H2})  we find the
form of the Rabi frequencies,
\beq
\tilde{\Omega}_{jk}=2e^{i(\phi_j-\phi_k)t}\Omega_{jk},
\eeq
with $\phi_1=0$,  $\phi_k$ ($k=2,3,4$) given by Eqs. (\ref{phases_1},\ref{Deltak}), and
$\tilde{\Omega}_{jk}=\tilde{\Omega}_{jk}+i\tilde{\Omega}'_{jk}$.

For other configurations that do not form a closed loop, similar steps may be followed,
but the four-photon resonance condition is not
imposed.

\end{document}